\documentclass[aps,pra,preprint,notitlepage,superscriptaddress,showkeys]{revtex4-2}

\usepackage[T1]{fontenc}
\usepackage{amsmath,amssymb}
\usepackage{graphicx}
\usepackage{url}

\begin{document}

\title{A theoretical study of subcarrier-wave quantum key distribution system integration with an optical transport network utilizing dense wavelength division multiplexing}

\author{F. Kiselev}
\email{kiselevfyodor@gmail.com}
\affiliation{Leading Research Center `National Center of Quantum Internet', ITMO University, Birzhevaya Line, 16, St. Petersburg, 199034, Russia}
\affiliation{Quanttelecom LLC., 6 Line, 59, St. Petersburg, 199178, Russia}

\author{N. Veselkova}
\affiliation{Leading Research Center `National Center of Quantum Internet', ITMO University, Birzhevaya Line, 16, St. Petersburg, 199034, Russia}

\author{R. Goncharov}
\affiliation{Leading Research Center `National Center of Quantum Internet', ITMO University, Birzhevaya Line, 16, St. Petersburg, 199034, Russia}

\author{V. Egorov}
\affiliation{Leading Research Center `National Center of Quantum Internet', ITMO University, Birzhevaya Line, 16, St. Petersburg, 199034, Russia}
\affiliation{Quanttelecom LLC., 6 Line, 59, St. Petersburg, 199178, Russia}

\begin{abstract}
In this paper we study the performance of subcarrier-wave quantum key distribution (SCW QKD) in the presence of classical channels of optical transport network utilizing dense wavelength division multiplexing (DWDM). We consider the impact of spontaneous Raman scattering noise as well as the four-wave mixing and channel isolation efficiency. We calculate quantum bit error rate as well as the secure key generation rate of SCW-QKD protocol for different parameters of DWDM system and quantum channel allocations. Our calculations show, that quantum channel allocation at the wavelength of 1310 nm is preferable to allocation at C-band due to lower nonlinear noise, despite fiber's larger attenuation at this wavelength.
\end{abstract}

\keywords{quantum communications, quantum key distribution, Raman scattering, four-wave mixing}

\maketitle

\renewcommand{\thetable}{\arabic{table}}

\section{Introduction}

Quantum key distribution (QKD) has become one of the most attractive communication technologies as it provides key distribution between two remote parties via open channel, security of which is guaranteed by the principles of quantum mechanics \cite{Gisin2002}. Usually, QKD requires dedicated optical fiber (or dark fiber), free of any other optical signals, to transmit quantum states without significant excess noise. This approach is often not feasible, due to continuously growing traffic demand and large costs of fiber deployment. In these circumstances, natural solution would be to integrate QKD system with currently deployed optical transport networks in a way that would allow for quantum and classical information channels to propagate within the same optical fiber. As it was shown in \cite{Dynes2016,Mao2018}, it is possible with wavelength-division multiplexing technology.

In recent years, dense wavelength division multiplexing (DWDM) has become widely adapted as a main transport method for high-capacity optical fiber communication systems \cite{Bahrani2018,Frohlich2017,Kumar2015,Niu2018}. Multiplexing QKD with classical optical channels over DWDM systems is challenging owing to the dense frequency grid and the large number of classical channels shared. Multiple noise suppression techniques have been suggested, such as the use of narrow-band filters \cite{Dynes2016,Eraerds2010,Frohlich2017,Mora2012,Patel2014,Wang2015}, temporal filtering technology \cite{Dynes2016,Patel2014,Wang2015}, as well as reducing the launch powers of classical optical signals \cite{Eraerds2010,Frohlich2017,Mora2012,Patel2014}. Since quantum signals are very weak compared to signals of classical optical channels, they are easily affected by impairments from classical light. The degradation of the quantum signals in DWDM-QKD system are caused mainly by two effects namely the four-wave mixing (FWM) noise \cite{Silva2014,Niu2018,Sun2016}, and the spontaneous Raman scattering (SpRS) noise \cite{Bahrani2018,Silva2014,Niu2018}. Such nonlinear effects are comparatively small in optical fibers transmitting a single optical channel. They become much more significant when multiple channels are packed into one fiber by wavelength division multiplexing.

For the first time, multiplexing of QKD channel with classical signals was presented by Townsend in 1997 \cite{Townsend1997}. In the proposed scheme the classical signals were located at the telecom C-band (1530--1565 nm), whereas the quantum signals were placed within the O-band (1260--1360 nm) through coarse wavelength-division multiplexing (CWDM) components. In this case, quantum signals experience less disturbances compared to the DWDM-QKD scheme since they are far from the classical signals spectral band. Later this CWDM-QKD scheme has been considered as a feasible solution and widely explored \cite{Aleksic2015,Chapuran2009,Choi2011,Wang2017}. However, in terms of deployment it could be more convenient to place both quantum and classical channels in the C-band using the DWDM technology due to the compatibility of such system with commercial optical network facilities and low fiber transmission loss in this spectral band. In the last decade, it is the DWDM-QKD scheme that has attracted more and more research \cite{Bahrani2018,Dynes2016,Eraerds2010,Frohlich2017,Kumar2015,Mora2012,Patel2014,Peters2009,Silva2014,Sun2016,Wang2015,Xavier2011}. In addition, an optimized channel allocation scheme was proposed \cite{Niu2018} for a DWDM-QKD system with multiple classical and quantum channels, which takes into account both FWM and Raman noise and is able to decrease the influence of these noises on the quantum signals.

In this paper, we have performed a theoretical research and numerical simulation of the noise influence caused by of SpRS, FWM and linear channel crosstalk on the performance of subcarrier-wave quantum key distribution (SCW QKD) system \cite{Kiselev:21,miroshnichenko2018security,Mora2012} integrated with classical DWDM channels. In SCW QKD system, the quantum signal is not directly emitted by the light source, but generated at the sideband frequencies via phase modulation of a strong monochromatic wave at the central frequency by use of an electro-optic modulator \cite{Miroshnichenko2017}; each quantum channel can be considered as a pair of subcarrier waves (or a single subcarrier \cite{Kiselev2020AnalysisSystem}). The advantages of these QKD systems are the absence of complex distributed interferometry schemes and the simplicity of phase shift matching in the transmitting and receiving modules. Besides that, SCW QKD systems demonstrate extensive capabilities of signal multiplexing. For example, the actual most advanced QKD systems demonstrate only 2\%--4\% spectral efficiency, while SCW systems can potentially enhance the spectral efficiency up to about 50\% \cite{Mora2012}.

The efficiency of the DWDM-QKD systems with different schemes of the channel allocation has been investigated for a multiplicity of sifting protocols, such as BB84 \cite{Eraerds2010,Mlejnek2017,Niu2018,Silva2014}, COW \cite{Mlejnek2017}, SARG \cite{Eraerds2010,Mlejnek2017}, and Gaussian-modulated coherent state protocol \cite{Kumar2015}. The QKD protocols utilizing the decoy-states \cite{Silva2014}, and `plug and play' phase encoding \cite{Eraerds2010} were also explored in the context of the WDM. In these works, such characteristics of the QKD channel as the quantum bit error rate (QBER) and the secure key generation rate have been obtained for multiple channel allocation schemes at different parameters of the system, namely the fiber length, the fiber attenuation, the launch power and the number of the classical channels, the receiver bandwidth, and the classical channel modulation parameters.

We analyze the impact of noise associated with the presence of classical DWDM channels in the same fiber with quantum channel of SCW QKD system supplementing our earlier work \cite{Kiselev:21}, which did not take into account FWM noise or linear channel cross-talk. We calculate secret key generation rate for systems of various configurations including placement of quantum channel in a uniform DWDM grid that occupies C-band of a telecommunication window and also configuration with allocation of quantum channel in O-band at 1310 nm. In general, in the case when the quantum channel is located in the C-band, we can conclude that FWM noise is mostly dependent on the channel spacing of the grid, while SpRS mainly depends on the number of channels and that for most configurations of DWDM grid SpRS is the dominant noise source for SCW QKD. Impact of FWM noise becomes significant for configurations with small number of channels and low spacing values.

Finally, we compare secret key generation rate of O-band configuration with C-band. With O-band configuration SCW QKD system shows significant improvement in terms of maximum achievable distance as compared to the C-band channel allocation. It was also shown that in order to achieve performance close to one shown in dark fiber at 1310 nm the extinction of quantum channel filtering system should be higher than 110 dB.

\section{Noise sources in DWDM}

\subsection{Spontaneous Raman scattering noise}

It is known that there are two types of SpRS noise depending on whether the signal is co-propagating with the pump light or counter-propagating with it. The first case is called forward SpRS noise and can be written as follows \cite{Eraerds2010,Mlejnek2017}:
\begin{equation}
P_{\mathrm{ram},f}=P_{\mathrm{out}}L\sum_{c=1}^{N_{\mathrm{ch}}}\rho(\lambda_c,\lambda_q)\Delta\lambda.
\label{ram_pw_f}
\end{equation}
The second case is called backward SpRS noise and can be written as:
\begin{equation}
P_{\mathrm{ram},b}=P_{\mathrm{out}}\frac{\sinh(\xi L)}{\xi}\sum_{c=1}^{N_{\mathrm{ch}}}\rho(\lambda_c,\lambda_q)\Delta\lambda,
\label{ram_pw_b}
\end{equation}
where $P_{\mathrm{out}}$ is the output power from the optical fiber for a single channel; $\xi$ is the attenuation of the fiber; $L$ is the length of the fiber; $N_{\mathrm{ch}}$ is the number of classical channels; $\rho(\lambda_c,\lambda_q)$ is the normalized scattering cross-section defined for the wavelengths of classical ($\lambda_c$) and quantum ($\lambda_q$) channels and is taken from \cite{Eraerds2010}; and $\Delta\lambda$ is the bandwidth of the filter used for isolation of quantum channel from the classical ones.

Channel wavelength in the DWDM grid can be calculated from its frequency which is described by ITU standard as follows:
\begin{equation}
\nu_n\,(\mathrm{THz})=191.6+N\nu_{\mathrm{spacing}},
\label{grid}
\end{equation}
where $N$ is the channel number and $\nu_{\mathrm{spacing}}$ is the grid spacing.

We use output power in these formulas here, instead of the input ones, to account for the sensitivity of the receiver ($R_x$) and the insertion losses of the system ($IL$). Using following relation for the output power we can automatically assign optimal channel power that will meet bit error rate (BER) requirements for the DWDM system
\begin{equation}
P_{\mathrm{out}}\,(\mathrm{dBm})=R_x\,(\mathrm{dBm})+IL\,(\mathrm{dB}).
\end{equation}
With SpRS noise power calculated we can now transform it to a photon detection probability to use in our QBER and secure key generation rate calculations
\begin{equation}
p_{\mathrm{ram},f/b}=\frac{P_{\mathrm{ram},f/b}}{hc/\lambda_q}\Delta t\eta_D\eta_B,
\label{ram_prob}
\end{equation}
where $\eta_D$ is the detector quantum efficiency; $\Delta t$ is the detector gating time; $\eta_B=10^{-0.1IL}$ describes the optical losses in the receiver's module; $h$ is the Planck constant; and $c$ is the speed of light.

The numerical simulations given in section 4.1 show that the Raman scattering noise is a dominant source noise for SCW QKD system if the quantum channel is located in a uniform DWDM grid. On the contrary, the impact of this noise on the quantum signal becomes negligible when the quantum channel is located in O-band at 1310 nm (see section 4.2).

\subsection{Four-wave mixing}

FWM is a type of optical Kerr effect which occurs when two or more light fields are co-propagating within the fiber. Nonlinear interaction between these fields result in the rise of a new field at a new frequency. In classical optical fiber communication systems with DWDM it results in a channel cross-talk. For the purpose of integrating QKD with DWDM infrastructure it is reasonable to place quantum channel at a wavelength which does not intersect with FWM products. However, if one decides to put a quantum channel in a uniform grid, noise coming from FWM products needs to be evaluated because it may enter into the QKD channel as an in-band noise and cannot be eliminated by filters. Let us assume that we have three pump channels with frequencies $f_i$, $f_j$, $f_k$, and that they generate a new frequency $f_i+f_j-f_k$. Peak power of (this new field) the FWM noise is provided by the expression \cite{Markowski:16,Niu2018}:
\begin{equation}
P_{ijk}=\frac{\eta D^2\gamma^2 P_iP_jP_k e^{-\xi L}}{9\xi^2}(1-e^{-\xi L})^2,
\label{fwm_power}
\end{equation}
where $P_i$, $P_j$, and $P_k$ are input power values of the pump channels; $\gamma$ is the nonlinearity coefficient; $D$ is the degeneracy factor which is set to six for the degenerate case when all three pump frequencies are different, and set to three when two of three frequencies are equal; $\eta$ is the efficiency of the FWM process which can be found as
\begin{equation}
\eta=\frac{\xi^2}{\xi^2+\Delta\beta^2}\left(1+\frac{4e^{-\xi L}\sin^2(\Delta\beta L/2)}{(1-e^{-\xi L})^2}\right),
\label{fwm_eff}
\end{equation}
where $\Delta\beta$ is the phase matching factor which can be expressed as
\begin{equation}
\Delta\beta=\frac{2\pi\lambda^2}{c}|f_i-f_k||f_j-f_k|\left[D_c+\frac{dD_c}{d\lambda}\left(\frac{\lambda^2}{c}\right)(|f_i-f_k|+|f_j-f_k|)\right],
\label{fwm_phase}
\end{equation}
where $\lambda$ is the wavelength of the FWM light; $D_c$ and $dD_c/d\lambda$ are the dispersion coefficient and slope respectively. Given that, the noise power coming from FWM can found as the sum of FWM products, frequency of which coincides with the frequency of a quantum channel
\begin{equation}
P_{\mathrm{FWM}}=\sum P_{ijk},\quad (\text{if } f_i+f_j-f_k=f_q).
\label{sum}
\end{equation}
Finally, the photon count probability caused by this noise can be found as
\begin{equation}
p_{\mathrm{FWM}}=\frac{P_{\mathrm{FWM}}}{hc/\lambda_q}\Delta t\eta_D\eta_B.
\label{fwm_prob}
\end{equation}
As will be demonstrated in section 4.1, the effect of FWM becomes significant for configurations with a small number of channels and low spacing values if quantum channel is inserted in a standard DWDM grid.

\subsection{Linear channel crosstalk}

Since quantum signal is very weak with respect to the classical signals and extremely sensitive to every source of noise we should also account for inefficiency of the filter that isolates quantum channel from the classical ones. We can calculate the power that leaks from the filter into quantum channel using the formula
\begin{equation}
P_{\mathrm{LCXT}}\,(\mathrm{dBm})=P_{\mathrm{out}}\,(\mathrm{dBm})-ISOL\,(\mathrm{dB}).
\end{equation}
Similarly to equations~(\ref{ram_prob}) and (\ref{fwm_prob}) we can calculate photon detection probability $p_{\mathrm{LCXT}}$ at the single-photon detector of the QKD system
\begin{equation}
p_{\mathrm{LCXT}}=\frac{P_{\mathrm{LCXT}}}{hc/\lambda_c}\Delta t\eta_D\eta_B.
\label{fwm_lcxt}
\end{equation}
In section 4.2 we will show that the dominant source of noise in O-band configuration is the channel crosstalk and that the extinction must be higher than 110 dB in order to achieve performance close to one demonstrated in dark fiber at 1310 nm.

\section{SCW QKD security}

\subsection{Quantum bit error rate}

As described above, the contribution of different noise sources of the fiber channel is described by the average photon fractions. In accordance with the theory of Mandel and Wolf \cite{Mandel1995} these can be taken into account in the expression of detector click probability as additional terms in the linear approximation \cite{miroshnichenko2018security}
\begin{equation}
P_{\mathrm{det}}(\varphi_A,\varphi_B)=\left(\eta_D\frac{n_{\mathrm{ph}}(\varphi_A,\varphi_B)}{T}+\gamma_{\mathrm{dark}}\right)\Delta t+p_{\mathrm{ram}}+p_{\mathrm{FWM}}+p_{\mathrm{LCXT}},
\label{eq:Pdet}
\end{equation}
where $n_{\mathrm{ph}}$ is the mean photon number depending on users' modulation indices and phases, $\gamma_{\mathrm{dark}}$ is the dark count rate.

According to \cite{miroshnichenko2018security}, the parameters of the binary symmetric error and erasure channel \cite{Cover2006} between Alice and Bob are described as
\begin{align}
E&=P_{\mathrm{det}}(0,\pi+\Delta\varphi),
\label{eq:E}\\
1-G-E&=P_{\mathrm{det}}(0,\Delta\varphi),
\label{eq:G}
\end{align}
where $G$ is the conditional probability of receiving an inconclusive measurement result; and $E$ is the conditional probability of an incorrect bit measurement, $\Delta\varphi$ is the phase offset due to imperfect synchronization system. This gives us the QBER, that provides the probability for Bob to receive an erroneous bit, as follows
\begin{equation}
Q=\frac{E}{1-G}=\frac{P_{\mathrm{det}}(0,\pi+\Delta\varphi)}{P_{\mathrm{det}}(0,\Delta\varphi)+P_{\mathrm{det}}(0,\pi+\Delta\varphi)}.
\label{eq:QBER}
\end{equation}
In our calculations, we shall use an approximate version of this expression from reference \cite{Gaidash2019d}
\begin{equation}
Q=\frac{2\mu\tau\eta(1-\vartheta)(1-\cos(\Delta\varphi))+\tau\vartheta\mu_0\eta+p_{\mathrm{dark}}+p_{\mathrm{ram}}+p_{\mathrm{FWM}}+p_{\mathrm{LCXT}}}{4\mu\tau\eta(1-\vartheta)+2\tau\vartheta\mu_0\eta+2p_{\mathrm{dark}}+2p_{\mathrm{ram}}+2p_{\mathrm{FWM}}+2p_{\mathrm{LCXT}}}.
\label{eq:QBER-approx}
\end{equation}
where losses are consolidated into $\eta=\eta_B\eta(L)\eta_D$ with $\eta(L)=10^{-0.1\xi L}$, $\mu_0$ is the mean photon number in carrier mode before the modulation, and $\mu$ is the mean photon number in sidebands depending on the modulation index $m$, $\vartheta$ is the spectral filter efficiency, and $\tau$ is the power fraction of quantum signal that fits within gating time of the detector and is assumed to be equal to 1 in this paper.

\subsection{Secret key generation rate}

To analyze the impact of various noise sources on security of the protocol, we consider the asymptotic case of the keys of infinite length. It is acceptable for this study, as the described optical effects do not play role in the finite-key analysis. This also assumes that the sent states are independent and identically distributed. Moreover, it slightly expands the security proof by adding constraints in the form of additional terms (see \cite{kozubov2019finite}). According to \cite{devetak2005distillation}, for one-way QKD protocols with independent identically distributed information carriers the secure key generation rate in the presence of collective attacks is lower bounded by the Devetak--Winter bound
\begin{equation}
K=v_S P_B\left[1-\operatorname{leak}_{EC}(Q)-\max_E\chi(A:E)\right],
\label{Keq}
\end{equation}
where $v_S$ is the repetition rate, which is 100 MHz for system considered in this paper, $P_B=(1-G)/N$ is the probability of successful state detection if basis is guessed correctly by Bob ($N$ is the number of bases), the amount of information disclosed by Alice during error correction is $\operatorname{leak}_{EC}(Q)\geq h(Q)$ is limited by the Shannon bound, where $h(x)$ is the binary entropy, and the last term is the Holevo information.

Further we assume that Eve is not affected by described channel noises and calculate the Holevo bound using the result of reference \cite{miroshnichenko2018security} derived for collective beam-splitting attack. According to \cite{miroshnichenko2018security}, the secure key generation rate~(\ref{Keq}) can be estimated as follows
\begin{equation}
K=\frac{(1-G)v_S}{2}\left[1-h(Q)-h\left(\frac{1-e^{-\mu_0m^2}}{2}\right)\right].
\end{equation}

Finally, we introduce maximum achievable distance, which is a metric, that shows maximum distance at which secret key can still be generated and is derived from secret key generation rate dependence on fiber length as shown in \cite{Kiselev:21}.

\section{Modeling results}

\subsection{Quantum channel replacing a single DWDM channel}
\label{subsec:qber1}

\begin{figure}[ht]
\centering
\begin{minipage}[t]{0.49\linewidth}
\centering
\includegraphics[width=\linewidth]{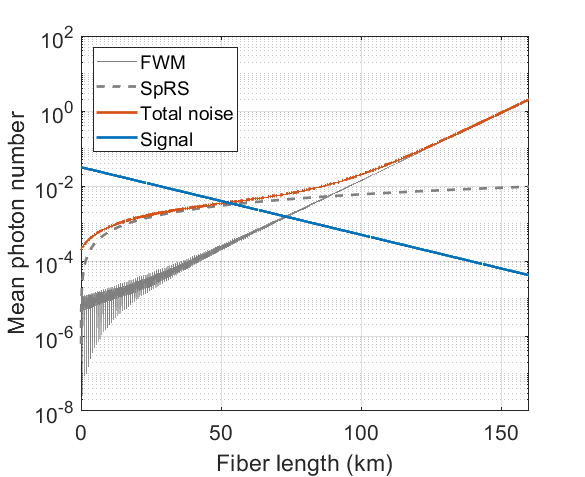}\\[-0.2em]
(a)
\end{minipage}\hfill
\begin{minipage}[t]{0.49\linewidth}
\centering
\includegraphics[width=\linewidth]{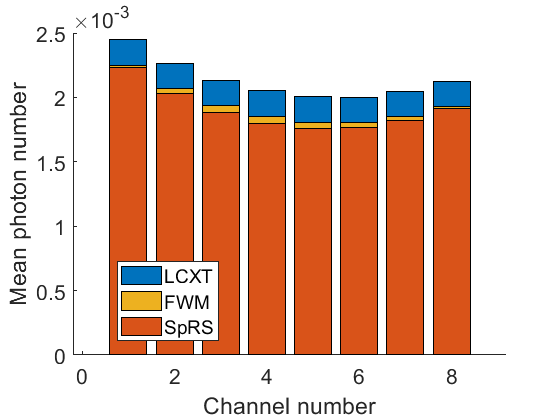}\\[-0.2em]
(b)
\end{minipage}
\caption{Simulation results for eight channel DWDM grid with 100 GHz spacing (a) mean photon number of different noise sources and quantum signal versus the fiber length; (b) mean photon number of different noise sources depending on the number of classical channel that is being replaced.}
\label{noise_1}
\end{figure}

First, we are going to take a look at the case, when we are replacing one of the channels in DWDM grid that occupies C-band of a telecommunication window. All fixed parameters of the system for our calculations can be found in table~\ref{tab1} in the appendix. This case was also considered in \cite{Patel2014} and is interesting to us, as it explores a possibility of a plug-and-play QKD-DWDM solution, where one does not have to make complex adjustments to DWDM optical scheme and can integrate QKD system by simply replacing the SFP modules dedicated to a single channel with an Alice and Bob modules of QKD system. Since in real life we are mostly dealing with equally spaced DWDM grids, FWM will become one of the noise sources that should be considered. First, let us take a look at the dependence of nonlinear noise mean photon number on the fiber length (figure~\ref{noise_1}(a)).

One can notice, that FWM noise grows much faster with the fiber length, completely overwhelming the SpRS noise at large distances. However, we are interested in the balance of noise components at the maximum achievable distance. Figure~\ref{noise_1} depicts the simulation results for the case of eight DWDM channels with 100 GHz spacing with one of the channels replaced with the quantum one. Maximum achievable distance for this case was calculated as 32 km. Figure~\ref{noise_1}(b) shows the balance between different noise sources depending on a channel number that is being replaced. One can see that at this distance noise from FWM has negligible impact in comparison with SpRS and even channel cross-talk.

\begin{figure}[ht]
\centering
\begin{minipage}[t]{0.49\linewidth}
\centering
\includegraphics[width=\linewidth]{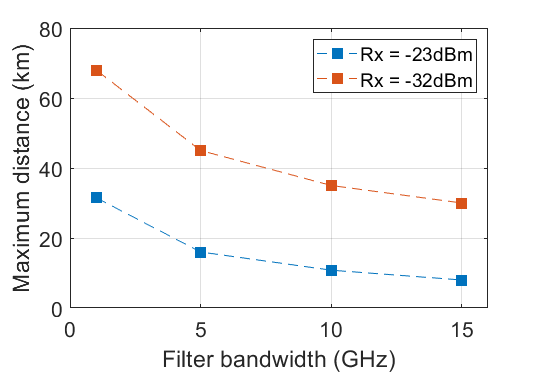}\\[-0.2em]
(a)
\end{minipage}\hfill
\begin{minipage}[t]{0.49\linewidth}
\centering
\includegraphics[width=\linewidth]{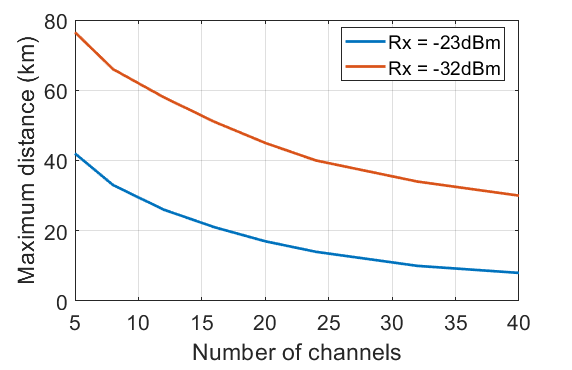}\\[-0.2em]
(b)
\end{minipage}
\caption{Simulation results for eight channel DWDM grid with 50 GHz spacing (a) photon detection probability due to noise and signal versus the fiber length; (b) photon detection probability due to different noise sources depending on the number of classical channel that is being replaced.}
\label{dist_dep}
\end{figure}

If we consider same amount of channels but mesh with lower spacing of 50 GHz, the maximum achievable distance will be decreased to 26 km due to increased power of FWM noise which can be seen on figure~\ref{noise_2}(b). Increasing the amount of channels to 40 will greatly increase noise levels due to SpRS and effect of FWM will again become negligible at the maximum achievable distance of 8 km (see figure~\ref{noise_3} in appendix). In general, it can be seen (figure~\ref{noise_dep}) that FWM noise is mostly dependent on the channel spacing of the grid, while SpRS mostly depends on number of channels. It also can be seen that for most configurations of DWDM grid SpRS noise is much larger than FWM. FWM becomes significant for smaller numbers of channels and grids with low spacing values. Finally, we can observe dependencies of maximum achievable distance versus filter bandwidth and number of channels. We can see figure~\ref{dist_dep}, that to achieve distances higher than 50 km one has to use smaller bandwidth and integrate QKD with low amount of classical channels.

\begin{figure}[ht]
\centering
\begin{minipage}[t]{0.49\linewidth}
\centering
\includegraphics[width=\linewidth]{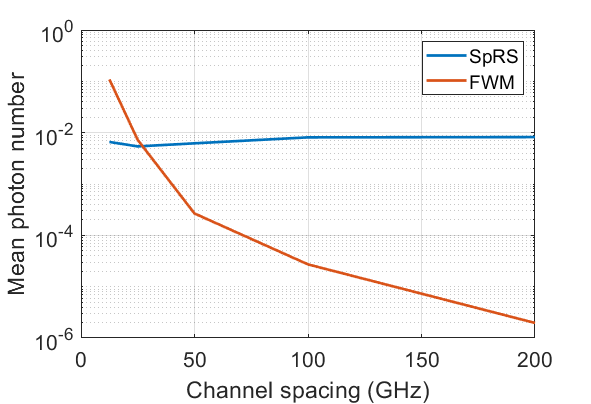}\\[-0.2em]
(a)
\end{minipage}\hfill
\begin{minipage}[t]{0.49\linewidth}
\centering
\includegraphics[width=\linewidth]{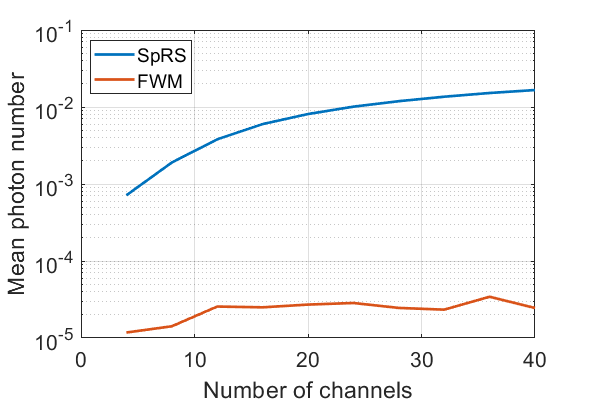}\\[-0.2em]
(b)
\end{minipage}
\caption{(a) Dependence of SpRS and FWM noise on channel spacing for 20 DWDM channels; (b) dependence of SpRS and FWM noise on number of channels for 100 GHz spacing.}
\label{noise_dep}
\end{figure}

\subsection{Quantum channel in O-band}
\label{subsec:qber2}

\begin{figure}[ht]
\centering
\includegraphics[width=0.8\linewidth]{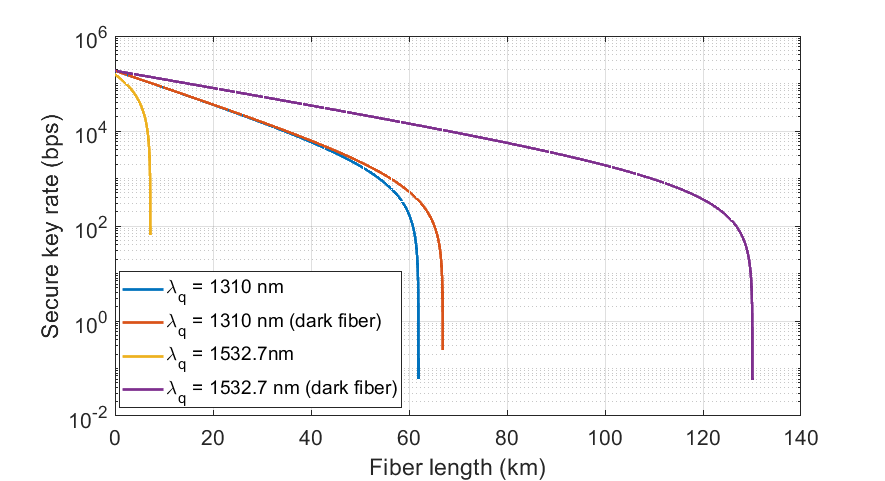}
\caption{Secure key generation rate versus optical fiber length for O-band and C-band QKD-DWDM configurations compared to cases with dark fiber. Fiber attenuation at 1310 nm is $\xi=0.35$ dB/km. Receiver sensitivity of information channels is $R_x=-23$ dBm.}
\label{reach_diff_bands}
\end{figure}

Another solution that seems to be get more success in recent years \cite{Wang2017} is the placement of quantum channel in the O-band of telecommunication window. By doing so we reduce the noise power of SpRS approximately by 4000 times \cite{Eraerds2010}. However, light experiences optical losses that almost twice as large in comparison with light at 1550 nm. This results in the fact that maximum achievable distance is approximately twice as smaller in comparison with the case when quantum channel is put in the C-band if we use dark fiber (meaning that there is no light present in the fiber other than from quantum channel). We calculated secret key generation rate versus optical fiber length for 40 channel DWDM grid with spacing of 100 GHz for four cases (figure~\ref{reach_diff_bands}): placement of quantum channel at 1532.7 nm wavelength in dark fiber and with DWDM, placement of quantum channel at 1310 nm wavelength in dark fiber and with DWDM.

\begin{figure}[ht]
\centering
\includegraphics[width=0.5\linewidth]{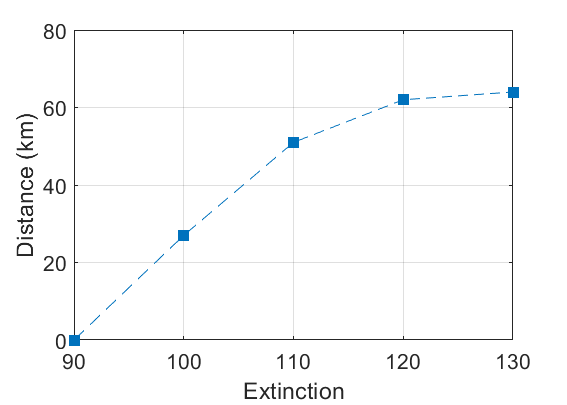}
\caption{Maximum achievable distance at which the SCW QKD system can operate versus extinction of filters of quantum channel.}
\label{dist_vs_ext}
\end{figure}

As one can see, despite higher attenuation, QKD system performs much better at 1310 nm in the presence of 40 classical channels. We found, that in this case largest noise contribution comes from LCXT, which was identified in \cite{Eraerds2010}. Figure~\ref{dist_vs_ext} depicts the dependence of maximum achievable distance versus filter's extinction in dB. One can see that in order to achieve performance close to one shown in dark fiber the extinction of filtering system of quantum channel should be higher than 110 dB.

We would like to note that presented analysis can be done for various QKD systems, including continuous-variable QKD schemes \cite{Chen2018,Laudenbach2018,Lupo2018}. In order to do so, noise calculations must be performed in the form of excess noise so it could be incorporated into the respective expression of covariance matrix.

\section{Conclusion}

We analyzed the impact of noise associated with the presence of classical DWDM channels in the same fiber with quantum channel of SCW QKD system. We calculated secret key rate generation for systems of various configurations including placement of quantum channel in a uniform DWDM grid and placement of quantum channel in O-band at 1310 nm. It was shown that SpRS is the dominant source noise for SCW QKD if quantum channel positioned in C-band. Impact of FWM becomes significant for configurations with low number of channels and spacing. Finally, we compared secret key generation rate of O-band configuration with C-band. O-band configuration shows significant improvement in terms of maximum achievable distance over C-band one. It was also shown that dominant source of noise in O-band configuration is LCXT and that it is required to have extinction higher than 110 dB in order to achieve performance close to one shown in dark fiber at 1310 nm.

\section*{Acknowledgments}

The work was done by Leading Research Center `National Center of Quantum Internet' of ITMO University during the implementation of the government support program, with the financial support of Ministry of Digital Development, Communications and Mass Media of the Russian Federation and RVC JSC; Grant Agreement ID: 0000000007119P190002, Agreement No. 006--20 dated 27.03.2020.

\section*{Data availability statement}

All data that support the findings of this study are included within the article (and any supplementary files).

\section*{Appendix}

\begin{figure}[ht]
\centering
\begin{minipage}[t]{0.49\linewidth}
\centering
\includegraphics[width=\linewidth]{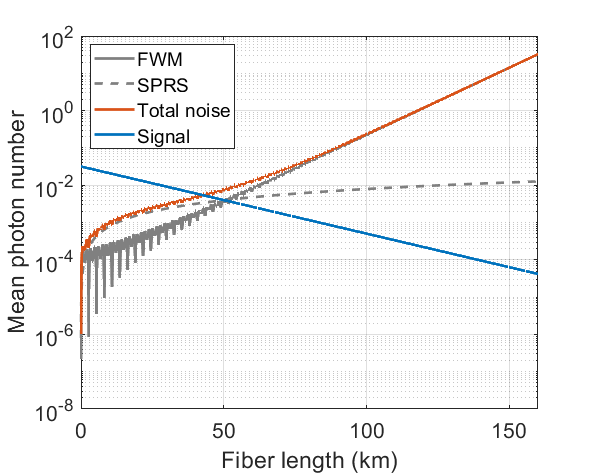}\\[-0.2em]
(a)
\end{minipage}\hfill
\begin{minipage}[t]{0.49\linewidth}
\centering
\includegraphics[width=\linewidth]{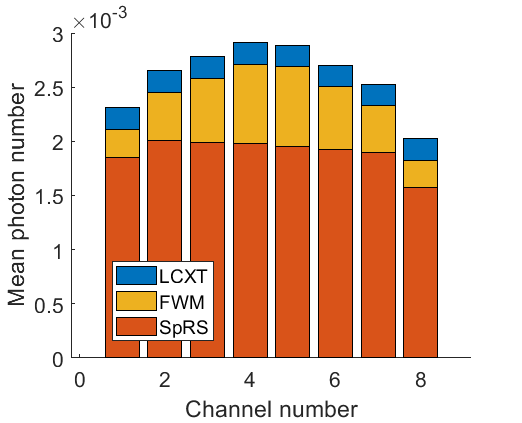}\\[-0.2em]
(b)
\end{minipage}
\caption{Simulation results for eight channel DWDM grid with 50 GHz spacing (a) mean photon number of different noise sources and quantum signal versus the fiber length; (b) mean photon number of different noise sources depending on the number of classical channel that is being replaced.}
\label{noise_2}
\end{figure}

\begin{figure}[ht]
\centering
\begin{minipage}[t]{0.49\linewidth}
\centering
\includegraphics[width=\linewidth]{fwm_sprs_8ch.png}\\[-0.2em]
(a)
\end{minipage}\hfill
\begin{minipage}[t]{0.49\linewidth}
\centering
\includegraphics[width=\linewidth]{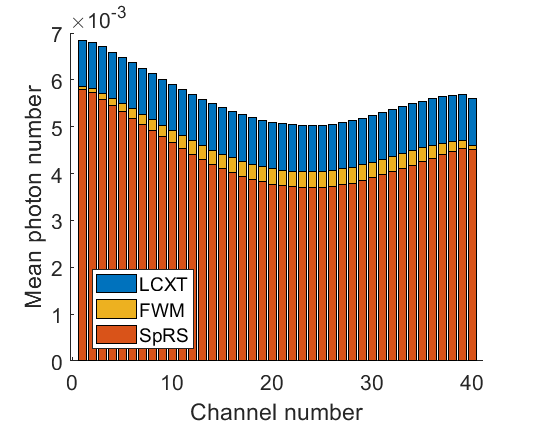}\\[-0.2em]
(b)
\end{minipage}
\caption{Simulation results for 40 channel DWDM grid with 50 GHz spacing (a) mean photon number of different noise sources and quantum signal versus the fiber length; (b) mean photon number of different noise sources depending on the number of classical channel that is being replaced.}
\label{noise_3}
\end{figure}

(See table~\ref{tab1}, figures~\ref{noise_2} and \ref{noise_3}).

\begin{table}[ht]
\caption{Parameters of optical transport network and QKD system.}
\label{tab1}
\centering
\begin{ruledtabular}
\begin{tabular}{lclc}
Parameter & Value & Parameter & Value \\
$\xi$ & 0.18 dB/km & $\mu_0$ & 3.93 \\
$D_\lambda$ & 16.0 ps $(\mathrm{nm}\times\mathrm{km})^{-1}$ & $m$ & 0.316 \\
$dD_\lambda/d\lambda$ & 0.09 ps $(\mathrm{nm}^2\times\mathrm{km})^{-1}$ & $p_{\mathrm{dark}}$ & $4\times10^{-6}$ \\
$R_x$ & $-23$ dBm & $\Delta\varphi$ & $5^\circ$ \\
$\Delta\lambda$ & 15 GHz & $\Omega$ & 4.8 GHz \\
$\eta_D$ & 0.1 & $\vartheta$ & $10^{-3}$ \\
$IL$ & 8 dB & $v_S$ & 100 MHz \\
$\gamma$ & 0.78 W$^{-1}$ km$^{-1}$ & $\Delta t$ & 1 ns
\end{tabular}
\end{ruledtabular}
\end{table}

\section*{ORCID iDs}

F. Kiselev \quad \url{https://orcid.org/0000-0002-3894-511X}

\bibliography{references}

\end{document}